Main Manuscript for

# Water adsorption on a model silicate surface: wollastonite (100)

Luca Lezuo,[a†] Andrea Conti,[a†] Alexander Hoheneder,[a] Elena Vaníčková,[b] Domitilla Alessandra Aloi,[a] Rainer Abart,[c] Florian Mittendorfer,[a] Michael Schmid,[a] Ulrike Diebold,[a] Giada Franceschi[a*]

[a]Institute of Applied Physics, TU Wien, 1040, Vienna, Austria

[b]Central European Institute of Technology, Brno University of Technology, 61200, Brno, Czech Republic

[c]Department of Lithospheric Research, Universität Wien, 1090, Vienna, Austria

† These authors contributed equally.

*Corresponding author: Giada Franceschi

**Email:** franceschi@iap.tuwien.ac.at

Water adsorption on silicate surfaces is a critical yet poorly understood process relevant to, e.g., mineral weathering and cement hydration. This study investigates the structure of water overlayers on a model calcium silicate, the lowest-energy (100) surface of wollastonite ($CaSiO_3$). It combines atomically resolved non-contact atomic force microscopy (nc-AFM), acquired with qPlus sensors and functionalized tips in ultrahigh vacuum (UHV), with density functional theory (DFT) calculations employing the metaGGA $r^2$SCAN+rVV10 functional. Adding incremental doses of water to the sample at cryogenic temperatures produces distinct structures governed by the competition between water–surface and water–water interactions. With two water molecules per surface unit cell, water–surface interactions dominate: In line with previous theoretical predictions, adsorbates follow the surface lattice. As the coverage increases, intermolecular hydrogen bonding competes with bonding to the surface, leading to the emergence of complex, coexisting patterns. While their small energy differences prevent an unambiguous identification of the most stable structure by DFT, the experimentally observed symmetries help constrain plausible structural models. Above a critical density of four water molecules per unit cell, water–water interactions prevail, and water clusters are formed. The results provide an atomic-scale framework for understanding water interactions with calcium silicate surfaces.

## Introduction

At the molecular scale, the behavior of water at solid interfaces is governed by a delicate balance between water–water cohesion and water–surface adhesion. This competition is central to processes as diverse as ice nucleation, surface wetting, catalysis, rock weathering, and cement hydration.[1–5] It determines whether water forms bulk-like, hydrogen-bonded (H-bonded) clusters or adsorption structures with a strong relationship to the underlying lattice.[6] The surface science approach can provide atomic-scale insights into this competition by studying water on idealized, single-crystalline surfaces in ultrahigh vacuum (UHV) and from cryogenic temperatures to room temperature. While this strategy has led to a deep understanding of water adsorption on metals[7–9] and selected oxide systems,[10–16] the interaction of water with natural silicate minerals remains less explored. Yet, water–silicate interactions underpin many key natural and technological processes, among others, the stabilization of concrete.[5]

The main binding material of concrete is cement, a hydrated paste rich in C–S–H (calcium-silicate-hydrate). C–S–H is a complex, heterogeneous structure built from "*dreierketten*" silicate chains (repeating units of three $SiO_4$ tetrahedra), interlayer calcium, and molecular water.[5] Because water actively participates in both the formation and the final structure of C–S–H, its role has been investigated extensively through simulations, including studies of C–S–H nucleation,[17] the distribution and reactivity of short silicate chains,[18] and the nature of interfacial water on C–S–H surfaces.[19] However, experimental benchmarks remain scarce: macroscopic observables such as the Ca:Si ratio, density, and mechanical properties provide only indirect information about the molecular-scale structure of C–S–H and its hydration. To address this limitation, well-defined crystalline analogues offer a promising route. Wollastonite ($CaSiO_3$), which contains corner-sharing silicate chains analogous to the *dreierketten* in C–S–H (see Fig. 1), is a structurally simple mineral platform for atomically resolved studies. Theoretical studies on wollastonite have addressed metal–proton exchange[20] and adsorption on low-energy surfaces.[21–23] Experimental insights have been limited to spectroscopic investigations of powders or polished aggregates,[24,25] primarily focusing on the (001) facet. Atomic-scale studies on water adsorption on well-defined surfaces, specifically the lowest-energy (100) surface, have been lacking.

Non-contact atomic force microscopy (nc-AFM) is a unique tool for filling the gaps in the experimental data. The development of qPlus sensors with piconewton force sensitivity[26] and of functionalized tips enabling sub-molecular resolution[27,28] provides the basis for the direct visualization of water adsorption at solid surfaces, including insulating materials.[29] Examples include water dimers, trimers, and tetramers on NaCl(001) thin films supported on Au(111) substrates,[30] 15-mer clusters on Pt(111) and Cu(111),[31] 2D bilayer hexagonal ice on Au(111),[32] water monolayers comprising 5–7-membered rings on Ni(111),[33] and 1D chains on Cu(110).[34] Combining precise water dosing in UHV with high-sensitivity force microscopy enables coverage-dependent measurements, which can directly probe computational predictions.

Such a combined approach is crucial, as modeling water structures at surfaces remains challenging. Density functional theory (DFT) predictions are highly sensitive to the choice of exchange-correlation functional. Standard generalized gradient approximation (GGA) functionals systematically underestimate van der Waals dispersion forces, which are crucial for water–surface binding.[35–38] In contrast, the metaGGA r$^2$SCAN functional, combined with the nonlocal rVV10 correlation,[39] provides improved descriptions that accurately capture both long-range physisorption and short-range interactions.[40] However, a fundamental discrepancy between theory and experiment often persists: DFT predicts ground-state structures at 0 K, whereas finite-temperature experiments reveal configurations of slightly higher energy.[16,41] This gap reflects entropic contributions and kinetic barriers that stabilize weakly bonded water species, particularly when entropy-driven effects prevail over small energy differences. Beyond these thermochemical effects, quantum effects of H nuclei can significantly influence dissociation barriers and network stability.[42,43]

In this study, the experimental and theoretical challenges are addressed by combining nc-AFM using a qPlus sensor in UHV with DFT calculations using the metaGGA r$^2$SCAN+rVV10 functional. The investigation focuses on water adsorption on the wollastonite (100) surface, the most abundant and environmentally relevant termination. Different adsorbate structures are resolved, characterized by (1 × 1), (√2 × √2)R45°, and (2 × 1) periodicities, and interpreted by DFT with support from AFM simulations.[44] The competition between water–water and water–surface interactions is crucial to the formation of the adsorption structures.

## Results

This section describes how water structures evolve on wollastonite (100) with increasing coverage: from isolated, strongly bound molecules that follow the lattice symmetry, to interconnected H-bonded networks with distinct symmetries, and finally to the formation of water clusters.

**1 $H_2O$/u.c.: nested water**

Fig. 1 illustrates the structure and appearance of the wollastonite (100) surface, as determined previously.[45] The crystal structure consists of chains of corner-sharing $SiO_4$ tetrahedra running along the [010] direction, separated by interchain Ca ions. These chains exhibit the afore-mentioned *dreierketten* configuration, the fundamental structural building block for C–S–H phases:[5] a corrugated repeating unit of three corner-sharing tetrahedra. Within this three-tetrahedra period, two (lower) paired tetrahedra point in one direction and a single bridging tetrahedron points in the opposite direction (upper tetrahedron). Upon cleaving, the surface exposes Ca ions arranged in a roughly rectangular unit cell (7.0 Å × 7.3 Å, see Fig. 1a). In nc-AFM images acquired with Cu-terminated tips, the Ca ions appear as dark features, as confirmed by the simulated AFM images based on the DFT model (Fig. 1b). Notably, the UHV-cleaved surface is already hydrated.[45] Water molecules released from the mineral sample during cleavage at RT readily adsorb in the valleys of the surface with a calculated adsorption energy of –1.3 eV/$H_2O$, resulting in a coverage of one $H_2O$ per unit cell. Each of these molecules (henceforth referred to as "nested") coordinates to two surface Ca ions via Ca–O bonds (≈2.53 Å) and donates a H-bond to a bridging oxygen linking the two lower $SiO_4$ tetrahedra of the *dreierketten*. While nc-AFM cannot identify the nested $H_2O$ molecules because they lie below the most protruding species, their existence is corroborated by the adsorption configuration of $CO_2$.[45] In DFT calculations, dissociated water species spontaneously recombined to form intact molecules (Fig. S6), a strong indication that molecular hydration is energetically favoured over dissociative adsorption. The cleaved surface naturally exhibits defects of currently unknown nature (Fig. 1c); this study focuses on defect-free regions.

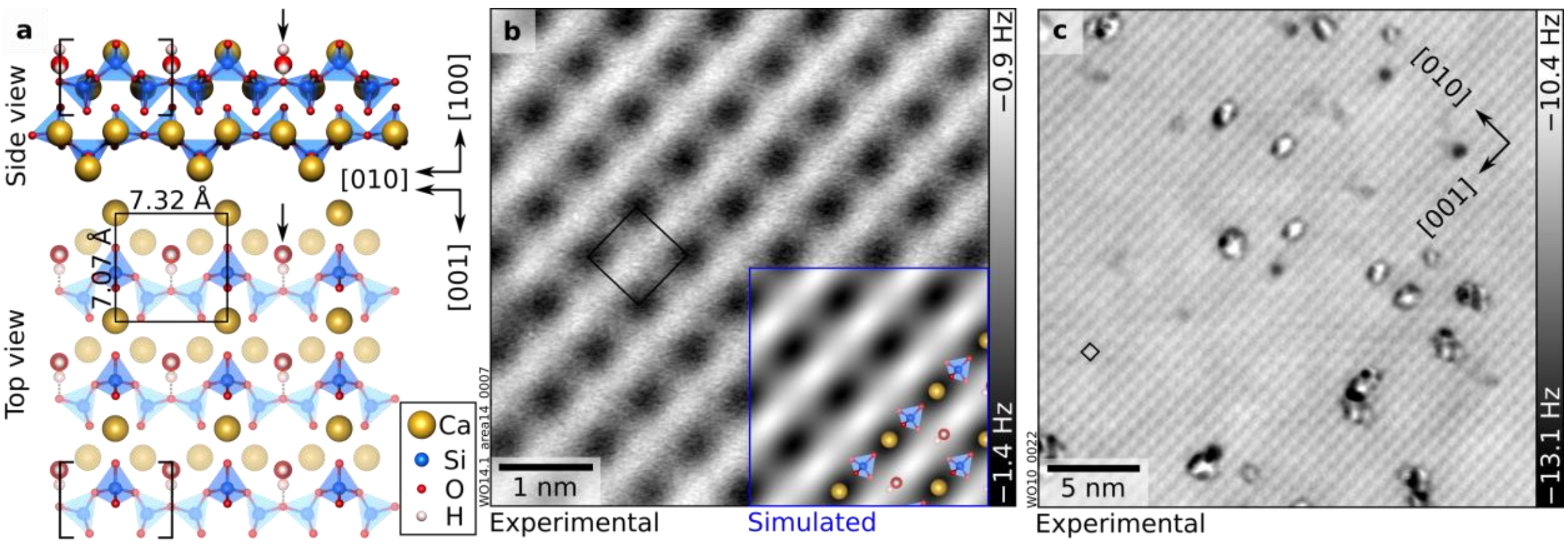

**Fig. 1. As-cleaved, hydrated wollastonite (100): One $H_2O$ molecule per unit cell.** (a) DFT-optimized surface structure of wollastonite (100) with one $H_2O$ per (1 × 1) surface unit cell adsorbed in the "nested" configuration (black arrow). Brackets indicate the *dreierketten* repeat unit (one upper and two lower silica tetrahedra) of the silicate chain running along the [010] direction. (b, c) 5.3 × 5.3 nm$^2$ and 27 × 27 nm$^2$ nc-AFM images of the UHV-cleaved surface that has reacted with $H_2O$ at room temperature. Dark grey indicates attractive tip-surface interaction, while in the light-gray regions, tip-sample interaction is weak. The inset in panel (b) shows an AFM simulation derived from the DFT model on the left-hand side (tip-sample distance of 5.7 Å). AFM parameters: (b) $A$ = 700 pm, $V_s$ = −2.5 V, Cu-terminated tip. (c) $A$ = 550 pm, $V_s$ = −10 V, wollastonite-modified tip. (1 × 1) unit cells are marked in black.

### 2 $H_2O$/u.c.: protruding water

Previous theoretical studies[21,22] have also predicted that water adsorption on wollastonite (100) is molecular, albeit with one molecule per unit cell adsorbing atop a Ca ion (with an adsorption energy of –0.9 eV/$H_2O$ as determined by empirical force-fields employed in ref.[22]). As noted above, the first $H_2O$ molecule per unit cell should instead adsorb in the "nested" configuration because of its stronger adsorption (Fig. 1). Upon placing further molecules on this hydrated surface (for a total of two $H_2O$ per (1 × 1) unit cell), the configuration predicted by Kundu *et al.* is found (Fig. 2a): $H_2O$ adsorbs atop Ca ions with one H atom forming an H-bond with a neighbouring O atom of the $SiO_4$ tetrahedron an adsorption energy of −0.97 eV/$H_2O$. The experimental data (Figs. 2b, c) confirm this assignment: Upon dosing water at 100 K, additional features become discernible in nc-AFM. Their coverage increases with the dose (Fig. S1), confirming their assignment to water species. At a coverage of roughly 0.45 ML (Fig. 2b), both the new protruding features assigned to water and the Ca ions of the cleaved surface are visible, confirming that protruding $H_2O$ and Ca ions are in registry. Furthermore, the simulated images based on the theoretical model closely match the experiments (inset in Fig. 2b; see Fig. S1 for a series of images as a function of tip–sample distance). The contrast is dominated by the H atom not engaged in bonding to the surface.

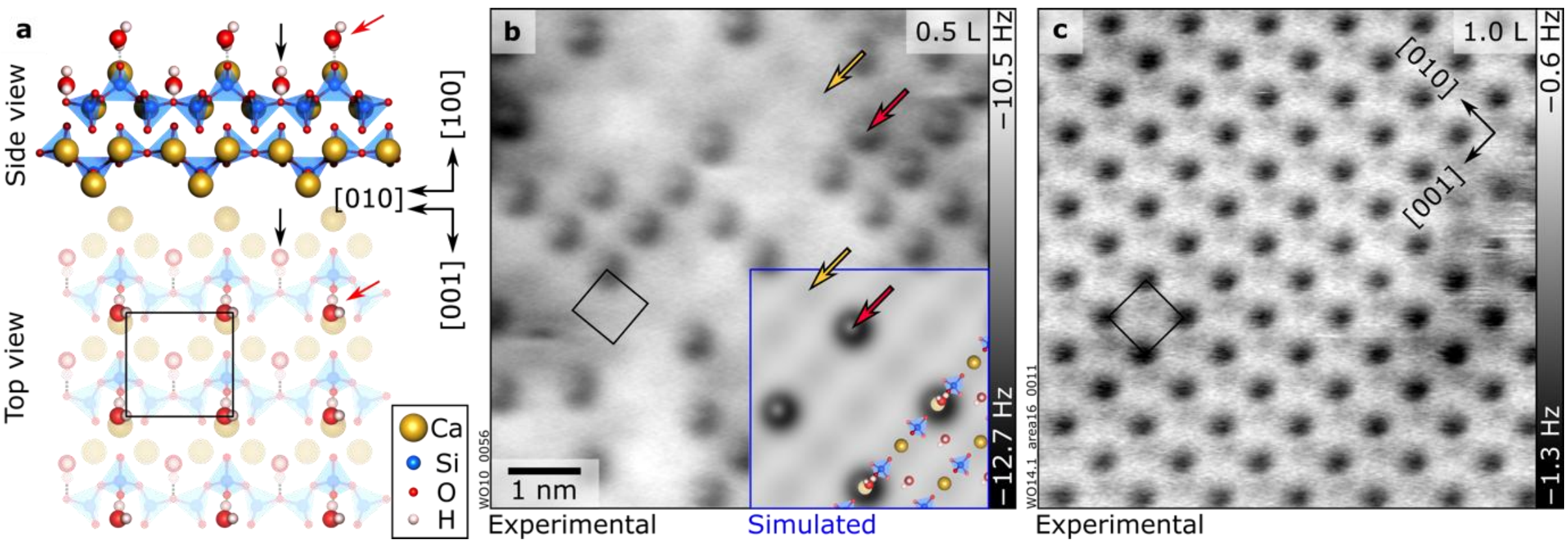

**Fig. 2: Two $H_2O$ molecules per unit cell: "protruding" $H_2O$.** (a) DFT-optimized surface structure of wollastonite (100) with one $H_2O$ in addition to the "nested" water, for a total of 2 $H_2O$ molecules per (1 × 1) unit cell. "Nested" and "protruding" water molecules are marked by black and red arrows, respectively. (b, c) 6.9 × 6.9 $nm^2$ nc-AFM images corresponding to coverages of "protruding" $H_2O$ species of 0.45 ML and 1 ML, respectively. Doses in Langmuir (L) of $H_2O$ dosed in addition to the nested water are indicated in the upper-right corners of the panels. AFM parameters: (b) $A$ = 200 pm, $V_s$ = −10 V; (c) $A$ = 700 pm, $V_s$ = −3.3 V; Cu-terminated tips. In panel (b), Ca atoms of the wollastonite lattice are faintly visible (yellow arrows) between the protruding water species (red arrows). The overlay in panel (b) shows an AFM simulation derived from the DFT model on the left-hand side (tip-sample distance of 4.5 Å). The AFM simulation was performed on a (2 × 2) supercell derived from the model in panel (a), with 0.25 ML of protruding water molecules per (1 × 1) unit cell (1.25 ML including nested $H_2O$). Black rectangles indicate the unit cell.

### 2.0-2.5 $H_2O$/u.c.: stripes and stable √2 patches

Beyond two molecules per unit cell (i.e., for coverages beyond the completion of a layer of protruding $H_2O$ with (1 × 1) periodicity as in Fig. 2c), two new patterns emerge (Fig. 3). Both have a local (√2 × √2)R45° symmetry and coexist with (1 × 1) patches of protruding $H_2O$. The blue and orange dashed rectangles in Fig. 3 highlight these new configurations, henceforth referred to as "stripes" and "stable √2" patches. Their contrast is affected by the tip termination (see Figs. 3a, b, acquired with O- and Cu-terminated tips, respectively, plus additional data in Figs. S2 and S3, which also show their relative alignment compared to the (1 × 1) protruding $H_2O$). Despite the same local periodicity, stripes and patches have distinct appearances in nc-AFM.

The stripes appear as isolated, meandering zigzag structures, often ending at defects (see Fig. S3d for a large-scale image with multiple stripes). They run along the [010] direction and are two (1 × 1) unit cells wide along the [001] direction. They tend to interact with the tip and cause the fuzzy appearance seen in Fig. 3. Sometimes the stripes move between consecutively taken images (Fig. S2), suggesting a relatively weak binding of species within the structure.

The "stable √2" patches are also elongated along [010] but are wider than the stripes along [001]. In contrast to the stripes, they are unperturbed by the tip. At low doses, the stripes predominate, and only a few small stable √2 patches are visible with some tips (Fig. S3d). As the coverage increases, the stripes become denser, although they stay mostly separated by one row of (1 × 1) protruding $H_2O$ features along [010] (cf. Figs. S3a, b). At the same time, larger stable √2 patches develop. Eventually, the stable √2 patches dominate the surface, until full coverage is achieved (Fig. 4). A statistical evaluation of nc-AFM images versus water doses reveals the same density of $H_2O$ per (1 × 1) unit cell in the two structures, namely 0.5 extra $H_2O$ per unit cell compared to the protruding $H_2O$ of Fig. 2c (corresponding to a total of 2.5 $H_2O$, including the nested $H_2O$). Although the coverage is now higher, in nc-AFM images, both structures display half the number of features compared to the full layer of protruding $H_2O$. As argued below, this is likely caused by the incorporation of the original 2 ML and water species into lower-lying configurations that are invisible to constant-height nc-AFM.

Fig. 4a shows a model for the stable √2 discussed in the next section. Using a (√2 × √2)R45° simulation cell, a satisfactory structural model for the stripes could not be identified. Under the hypothesis that the stripes might be a metastable structure, an annealing experiment (20 minutes between 120 K and 140 K) was performed on a surface with coexisting stripes and stable √2 patches. No significant changes were observed in the appearances and relative coverages of the structures. It is therefore likely that understanding the stripe structure would require a larger simulation cell, including the rows adjacent to the stripe. However, due to the significant computational effort required to probe many configurations in such a larger cell, this approach was not pursued.

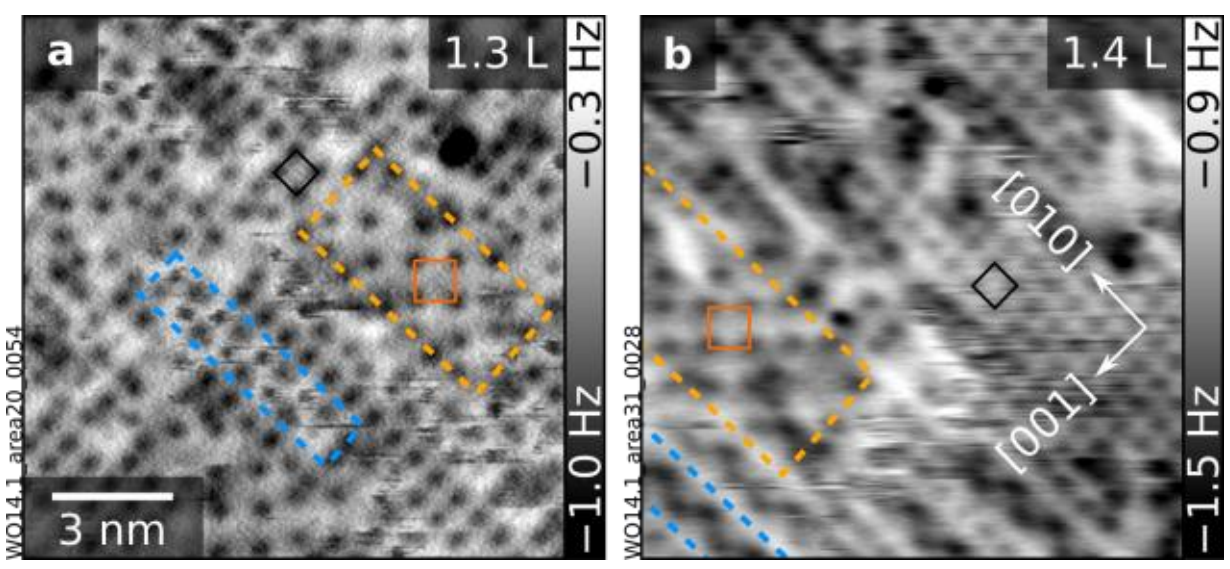

**Fig. 3: Between 2 and 2.5 $H_2O$ molecules per unit cell: "stripes" and "stable √2" patches.** (a, b) 13.5 × 13.5 nm$^2$ nc-AFM images taken with differently terminated tips. Areas with "stripes" and "stable √2" patches are outlined in blue and orange, respectively, and (1 × 1) and (√2 × √2)R45° unit cells are marked in black and dark orange, respectively. Doses in Langmuir are reported in the upper-right corner of the corresponding panels. AFM parameters: (a) $A$ = 700 pm, $V_s$ = −10 V, O-terminated tip; (b) $A$ = 700 pm, $V_s$ = −3.5 V, Cu-terminated tip.

**2.5 $H_2O$/u.c.: complete layer of "stable √2"**

At a total coverage of 2.5 molecules per (1 × 1) unit cell, an almost complete "stable √2" layer with antiphase domains is observed (Fig. 4b; domain boundaries are highlighted in Fig. S5). As noted above, the density of species detected by nc-AFM is smaller compared to the "protruding $H_2O$", despite the larger $H_2O$ doses needed to achieve this structure, hence suggesting the presence of lower-lying species that are invisible to nc-AFM in constant-height mode. To establish a structural model of the stable √2 pattern through DFT, 500 candidates were screened (see Computational Methods). Several energetically competing models were found, within an energy window of 0.1 eV per unit cell (see Fig.

S7 for exemplary configurations and corresponding energies). Such energy differences are too small to be significant, and the inaccessibility of lower-lying species in constant-height nc-AFM prevents a conclusive experimental validation. Nonetheless, most models show common features, with only minor deviations in the orientation of some H atoms (see Fig. S7). Fig. 4a illustrates the lowest-energy structure found. It suggests a cooperative reorganization of the hydration layer. The extra $H_2O$ added to the "protruding" configuration induces a structural reorganization: Every second protruding $H_2O$ moves to a lower position where it accepts one H-bond from the new $H_2O$ and donates one to the apical O atom of the surface $SiO_4$ tetrahedron (purple arrow). At the same time, the new $H_2O$ (green arrow) accepts an H-bond from the nested $H_2O$. The presence of multiple, lower-lying species explains the lower number of species observed in nc-AFM: The contrast is dominated by the remaining protruding $H_2O$ (Fig. 4b). While it is impossible to confirm the arrangement of low-lying species, AFM image simulations based on this proposed model are consistent with experiment.

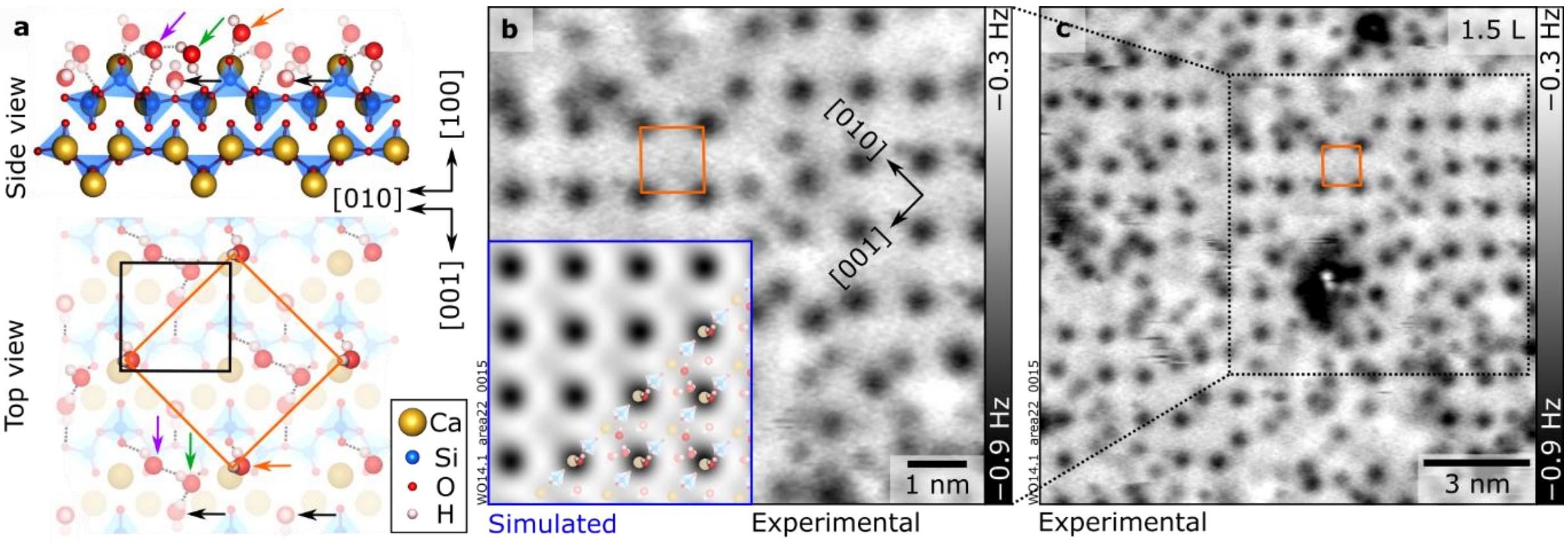

**Fig. 4: 2.5 $H_2O$ molecules per unit cell: Full layer of "stable √2".** (a) DFT-optimized surface structure of wollastonite (100) with a (√2 × √2)R45° overlayer containing 2.5 $H_2O$ per (1 × 1) unit cell (see Fig. S7a for a larger-scale top view of the model). The H-bonded network (marked by gray dotted lines) involves three $H_2O$ molecules: The nested one (black arrow), one previously protruding (purple arrow), and one additional $H_2O$ (green arrow). The remaining protruding molecules (orange arrow) are aligned with the surface Ca ions and dominate the AFM contrast (circular dark features). (b, c) nc-AFM images (8.4 × 8.4 $nm^2$ and 14 × 14 $nm^2$, respectively) corresponding to the same dose of 1.5 L ($A$ = 700 pm, $V_s$ = −2.5 V, O-terminated tip). The inset in panel (b) shows an AFM simulation derived from the DFT model on the left-hand side (tip-sample distance of 5.2 Å).

**2.5-4 $H_2O$/u.c.: (2 × 1)**

Upon dosing further water on the complete "stable √2" layer, a new configuration with (2 × 1) periodicity emerges (see Fig. 5 and Fig. S5 for the transition between the two structures). The new (2 × 1) structure was built from four $H_2O$ per (1 × 1) unit cell, as suggested by the experimental doses. It is characterized by [010]-oriented rows containing one dark, circular feature and one fainter, elongated feature per (2 × 1) unit cell (see the close-up image in Fig. 5e). Similar to the "stable √2" modeling, several of the candidates screened in DFT were found to be structurally and energetically close (Fig. S7). Fig. 5f shows the best model. The most protruding $H_2O$ (magenta arrow) donates an H-bond to the apical O atom of an upper surface $SiO_4$ tetrahedron, while simultaneously accepting H-bonds from two neighboring molecules. The first of these neighbors (blue arrow) accepts an H-bond from a nested $H_2O$ and donates an H-bond to another O of the upper $SiO_4$ tetrahedron. The second neighbor (orange arrow) is involved in an extended H-bonding network that involves lower Si tetrahedra of the *dreierketten* and includes the second-most protruding $H_2O$ species. This species accepts a bond from a nested $H_2O$ and donates to an apical O of the upper tetrahedron of the next row of *dreierketten*. Notably, in all low-energy structures, the nested $H_2O$ molecules remain rigidly bound to their original sites, underscoring their exceptional stability. This arrangement rationalizes the experimental contrast with (2 × 1) periodicity (see Fig. 5e, overlaid to the AFM simulation of the proposed model): The most protruding $H_2O$ molecule in each (2 × 1) unit cell (purple arrow) appears as a circular, prominent dark feature, while the molecule marked by a black arrow is seen as a fainter, elongated feature. Water dosed to the (2 × 1) structure does not trigger the formation of any new surface structure. Instead, three-dimensional clusters form on top of the (2 × 1)-ordered surface (see Fig. 5c).

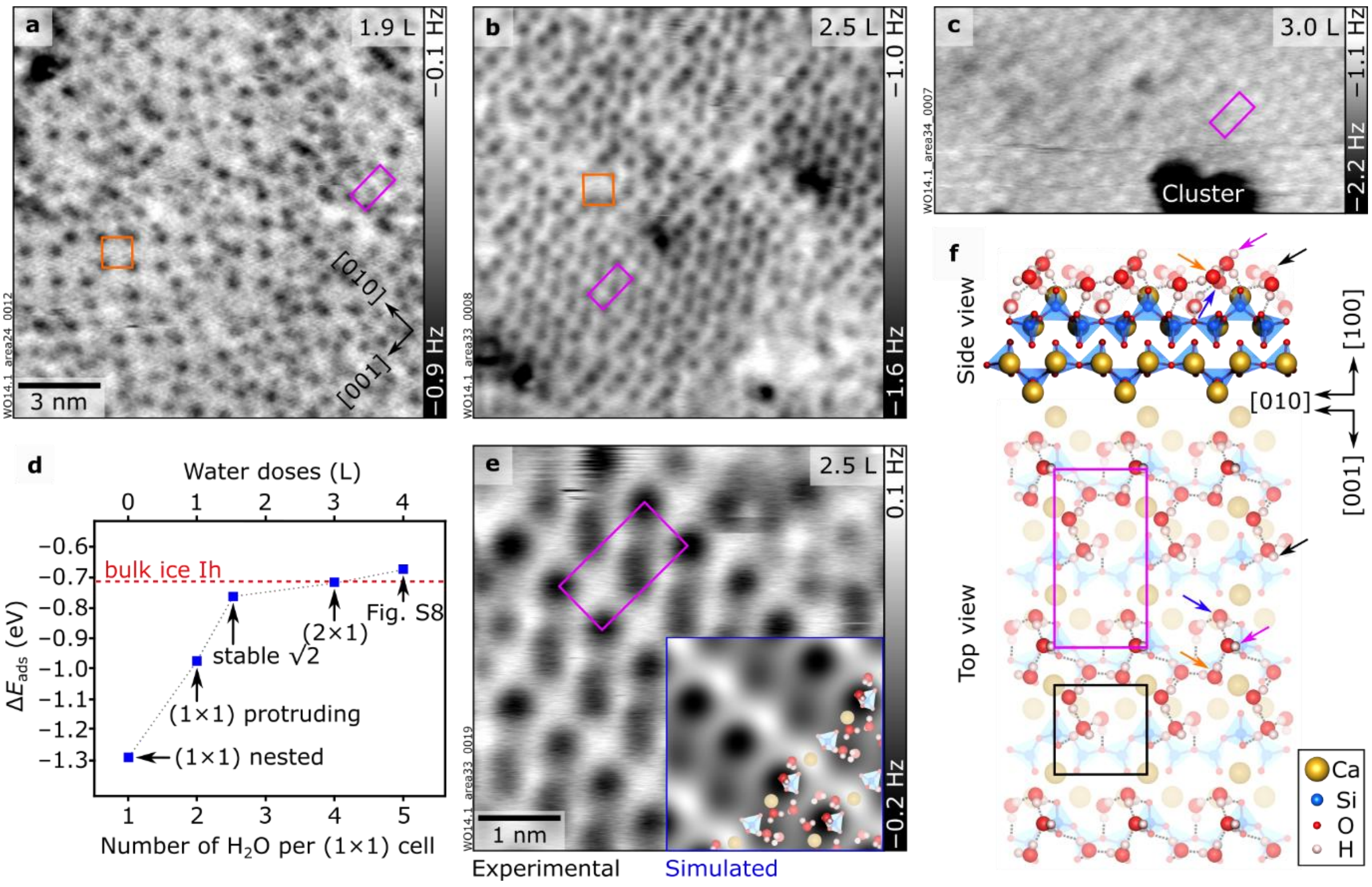

**Fig. 5: Between 2.5 and 4 $H_2O$ molecules per unit cell: from "stable √2" to (2 × 1) configurations.** (a, b) 15 × 15 $nm^2$ nc-AFM images of wollastonite (100) with increasing $H_2O$ coverages, moving from (a) the "stable √2" (orange square) plus additional features to (b) a (2 × 1) overlayer. Water doses in Langmuir (in addition to the nested water) are indicated in the top-right corners of the AFM images. (c) 14.7 × 7.6 $nm^2$ nc-AFM image showing a water cluster (dark; attractive interaction with the tip) on top of the (2 × 1) layer. (d) Differential adsorption energies ($\Delta E_{ads}$) of the calculated configurations at different $H_2O$ coverages. The dashed red line indicates the bonding energy of hexagonal bulk ice calculated with the same functional; the dotted, grey line connects the blue dots to visualize the trend. The blue markers correspond to the structure models of "nested" $H_2O$ (Fig. 1a), "protruding" $H_2O$ (Fig. 2a), full "stable √2" pattern (Fig. 4a), (2 × 1) pattern (panel f), and the lowest-energy model obtained for a coverage of five $H_2O$ per (1 × 1) unit cell (Fig. S8). (e) 5 × 5 $nm^2$ nc-AFM image of the (2 × 1) structure overlaid by the AFM simulation corresponding to the model in panel (f). AFM parameters: (a) $A$ = 700 pm; $V_s$ = –3.6 V; (b) $A$ = 700 pm; $V_s$ = –0.1 V; (c) $A$ = 700 pm; $V_s$ = –3.4 V (O-terminated tips). Orange squares and magenta rectangles identify the unit cells of the "stable √2" and (2 × 1) structures, respectively. (f) DFT-optimized structure model of wollastonite (100) with 4 $H_2O$ per (1 × 1) unit cell, forming a (2 × 1) pattern (see Fig. S7d for a larger-scale top view of the model). Magenta and black arrows indicate the most and second-most protruding $H_2O$ molecules, respectively. Blue and orange arrows mark the two neighboring molecules donating H bonds to the most protruding $H_2O$.

## Discussion

The literature on water–oxide interactions highlights their inherent complexity, as varied cation coordination on oxide surfaces, structural defects, and local electrostatic fields result in a broad spectrum of adsorption structures. For instance, water on rutile $TiO_2$(110) forms H-bonded dimers on hydroxylated surfaces;[10] MgO(001) develops ordered monolayer structures with partial dissociation;[11] CaO(001) exhibits partial dissociation and hydroxylation, forming ordered chains and networks at room temperature.[12,13] Half-dissociated monolayers coexist with molecular configurations on ZnO(10–10).[14] At initial coverages, $In_2O_3$(111) develops ordered hydroxyl overlayers,[15] while on $Fe_3O_4$(001), partially dissociated dimers act as anchors for extended H-bonded networks.[16] The intricacy demonstrated by these studies is further amplified in more complex minerals. On calcite ($CaCO_3$), water preferentially binds to carbonate group rows on the (2 × 1)-reconstructed surface, forming islands between 0.5 and 1 ML coverage that integrate into a (1 × 1) monolayer structure at full coverage.[63] Some silicate minerals, such as microcline feldspar ($KAlSi_3O_8$), undergo spontaneous hydroxylation at room temperature, in turn facilitating ordered water adsorption.[48] The formation of specific patterns on each system is governed by the complex interplay between water–water and water–surface interactions, which is modulated by the atomic details of the surface structure.

The wollastonite $CaSiO_3$(100) surface provides an opportunity to disentangle these contributions experimentally. At the lowest coverages, water molecules adsorb molecularly and occupy well-defined Ca sites, forming (1 × 1) motifs, characterized by "nested" and "protruding" water (Figs. 1 and 2). The unusually large adsorption energies calculated for 1–2 $H_2O$ per unit cell (–1.3 eV and –0.97 eV, respectively) indicate that electrostatic interactions with surface $Ca^{2+}$ dominate in this regime. The nested water molecules maximizes direct contact with the cations and act as anchors for subsequent adsorption, in close analogy to hydroxyl groups serving as nucleation points for H-bonded networks on other oxides.[15] Importantly, no evidence for water dissociation is found at any coverage, in agreement with earlier theoretical predictions[21,22] but distinct from the behavior of the (001) termination.[23–25] As recently demonstrated by direct atomic-scale measurements on model oxides, different surface oxygen sites can exhibit distinct proton affinities, which govern the surface's tendency to accept or abstract a proton from an adsorbate.[64] On the wollastonite (100) surface, the relatively low proton affinity (i.e., limited Brønsted basicity) of the surface oxygen atoms disfavors proton transfer from water and thus suppresses dissociative adsorption.

Above 2 ML (nested and protruding water adsorbed), more complex interactions and patterns emerge. The new molecules do not simply decorate every equivalent site, like at lower coverages where water–surface interactions dominate. The added $H_2O$ interacts with the nested $H_2O$ such that translational symmetry of the lattice is broken. Water–water interactions become more important in the new hydration patterns ("stable √2" and "stripes", see Figs. 3 and 4), ultimately giving rise to denser (2 × 1) arrangements and, beyond ≈4 $H_2O$ per unit cell, to three-dimensional clusters (Fig. 5c).

The differential adsorption energy for the addition of the fifth $H_2O$ molecule per unit cell (see Fig. S7) is –0.68 eV, which is less stable than the cohesive energy of bulk ice Ih (–0.71 eV; see Fig. 5d). This energetic crossover demonstrates that the system has shifted from a surface-templated regime into one governed by water–water cohesion, providing a thermodynamic rationale for the emergence of three-dimensional clusters. The onset of three-dimensional growth rather than the formation of extended ice-like overlayers can be understood by considering both water density and lattice symmetry. At the cluster-nucleation threshold, the surface water density is well below that of the ice Ih basal plane (see Table T2). Moreover, the wollastonite (100) surface imposes a rectangular lattice constraint that is incompatible with the hexagonal symmetry of bulk ice Ih, the thermodynamically stable crystalline phase at the investigated low-pressure conditions. This fundamental mismatch hinders epitaxial ice growth, in contrast, e.g., to the facile epitaxy supported by the AgI(0001) surface.[60]

The complexity of water adsorption on wollastonite (100) is underscored by the multivalley energetic landscape revealed by DFT: a multitude of distinct, competing configurations exist at coverages beyond two $H_2O$/u.c. (Fig. S7). These structures lie within a narrow energy window (<0.1 eV/u.c.) that falls within the typical accuracy limits of DFT for H-bonded systems.[38] This energetic near-degeneracy makes it challenging to pinpoint a unique ground-state structure. Instead, it points toward a family of structurally similar and energetically close configurations. This theoretical finding mirrors the coexisting structural motifs observed experimentally as well as the absence of sharp phase transitions, reflecting the gap between static predictions and finite-temperature experiments where entropic contributions and kinetic barriers stabilize weakly bonded species.[16,41]

## Conclusions

This study reveals the structural evolution of water adsorbed on a model calcium silicate surface, wollastonite $CaSiO_3$(100). A combination of atomically resolved nc-AFM and DFT demonstrates that water adsorbs molecularly with distinct motifs governed by the coverage-dependent competition between water–surface and water–water interactions.

At low coverage (one to two $H_2O$ per unit cell), the adsorption geometry is strictly governed by water–surface interactions, characterized by molecules anchored via electrostatic forces to surface calcium ions and hydrogen bonds to the silicate framework. In the intermediate regime (two to four $H_2O$ per unit cell), the increased water density triggers a competition between adhesion and intermolecular cohesion, resulting in a flat potential energy landscape populated by coexisting, complex hydrogen-bonded networks. Finally, at coverages exceeding four molecules per unit cell, water–water cohesion overcomes surface templating effects, leading to the nucleation of three-dimensional clusters.

The results provide mechanistic insight into how water interacts with calcium silicate surfaces, establishing a fundamental framework for understanding processes such as cement hydration and aqueous weathering. The near-degeneracy in energy of the coexisting surface phases poses a challenge for theoretical simulations. In addition, nc-AFM solely probes the uppermost, protruding water molecules rather than the full adsorption geometry. Nevertheless, the symmetry revealed by the experiments constrain the range of viable configurations and guide the computational analysis toward the establishment of structural models. Overall, this work demonstrates a robust symmetry-guided strategy for unraveling complex oxide–water interfaces.

## Methods

**Sample and ex-situ characterization.** A crystalline sample of wollastonite 1A was extracted from a larger specimen obtained from a skarn occurrence in Turkey. The specimen was previously characterized and analyzed by light and polarization microscopy, electron probe microanalysis, in-air AFM, X-ray diffraction, and electron back-scattered diffraction (EBSD).[45] Throughout this work, we adopted the conventional unit cell notation (bulk lattice parameters should fulfill $a > b > c$), which is standard in mineralogical databases and other works[21,22,45–47] and leads to the labeling of the lowest-energy surface as (100). The (100)-oriented grains (see EBSD analysis in Ref.[45] for the orientation determination) were glued onto Omicron-style stainless-steel plates using a UHV-compatible epoxy glue (EPO-TEK T7110-38), and a metal stud was glued to the top of the sample. A tangential force applied to the stud cleaved off the portion of the sample initially covered by the stud.[48] During cleaving, the pressure in the UHV chamber spiked by a few orders of magnitude, as previously reported for this and other silicate minerals.[45,48] An optical microscope attached to the UHV chamber helped identify portions of the cleaved sample that were flat enough to approach the nc-AFM tip.

**UHV setup.** The experiments were carried out in a UHV setup consisting of two interconnected chambers: a preparation chamber for sample cleaving, gas dosing at 100 K, and X-ray photoelectron spectroscopy (XPS) with a base pressure below $1 \times 10^{-10}$ mbar, and an adjacent chamber for nc-AFM ($1 \times 10^{-11}$ mbar). Ultrapure water (MilliQ, Millipore, 18.2 MΩ cm, ≤ 3 ppb total organic carbon) was dosed into the preparation chamber through a leak valve while keeping the sample at 100 K. The water was purified through several freeze-pump-thaw cycles.[49] The amount of water deposited was estimated based on the dose (partial pressure × time) required to achieve a coverage of one molecule per unit cell (u.c.), under the assumption of 100% sticking probability at 100 K, and verified by counting features on the surface by analyzing the nc-AFM images. These calculations yielded an approximate coverage of 1 $H_2O$ per unit cell (u.c.) per 1 Langmuir (L) dosed on top of the as-cleaved surface (1 L = $1.33 \times 10^{-6}$ mbar · s). Nc-AFM confirmed that all dosed water molecules desorbed from the surface after warming up to room temperature. One monolayer (ML) is defined as one molecule per (1 × 1) unit cell.

**Nc-AFM.** The nc-AFM measurements were performed in constant-height mode at 78 K, using a commercial Omicron qPlus low-temperature (LT) head and a differential cryogenic amplifier.[50] The qPlus AFM sensors ($k$ = 2,000−3,500 N/m, $f_0 \approx$ 27 or 32 kHz, $Q \approx$ 14,000 or 12,000) with a separate contact for the tunneling current had etched W wire tips glued to the oscillating prong of the sensor. In the UHV chamber, the tip was prepared in STM mode on a partially oxidized Cu(110) surface.[28] Soft indentation and voltage pulses were repeated until atomically sharp tips with a frequency shift of less than −1.5 Hz were achieved at the tunneling settings of 500 mV and 200 pA. The tips were terminated with Cu or O as judged by the contrast in constant-current STM.[45] This enabled chemically sensitive AFM and comparison to simulated images derived from density functional theory (see Computational Methods). Note that the constant-height mode makes AFM sensitive only to the most protruding species.

To mitigate surface charges developed during UHV cleaving, the sample was irradiated for one minute with X-rays from the XPS setup.[48] Residual charges were assessed using the Kelvin parabola method[51] and compensated by applying a bias voltage, $V_s$, to the back of the sample plate while keeping the tip potential close to ground.

**Data analysis.** Raw images were processed in ImageJ[52] to correct distortions arising from piezo creep and thermal drift[53] and to remove low-frequency noise. Limited sampling areas, tip-dependent imaging contrast (Figs. 3 and S4), and uncertainties in the dosing procedure may introduce errors in the estimated water densities for each structure. Nevertheless, the estimates are considered reliable, as experiments performed with incremental dosing yielded results consistent with those obtained using a single, larger dose.

**Computational methods**

This study was conducted using standard density functional theory (DFT) calculations. Although machine-learned force fields (MLFF) can represent a computationally effective alternative, the computational cost of the on-the-fly approach implemented in VASP scales quadratically with the number of atomic species.[54] This unfavorable scaling renders the method inefficient for the wollastonite system, which contains five species: Ca, Si, O-wollastonite, O-water, and H. In addition, accurately modeling the oxide–water interface including the requisite van der Waals forces, is challenging for MLFF.[55] Since the (1 × 1) surface unit cell (in-plane dimensions: ≈ 7.07 Å × 7.32 Å) was computationally tractable, standard DFT was the more reliable and preferred method.

**Calculation details.** DFT calculations were performed using the projector-augmented wave (PAW) method,[56,57] as implemented in the Vienna Ab-initio Simulation Package (VASP, version 6.5.1).[58] The metaGGA $r^2$SCAN+rVV10[36,37,39] exchange-correlation functional was employed, unless otherwise noted. These functionals were chosen for their good description of the bulk properties of wollastonite, with lattice parameters and angles deviating by less than 0.5% from the values measured by XRD (see Table T1). The bulk structure was optimized using a cutoff energy of 800 eV, a 3 × 3 × 3 k-point mesh for Brillouin zone integration, and a Gaussian smearing with a width of 0.1 eV. For the (100) surface calculations, symmetric slabs were constructed. They had a thickness of three bulk unit cells (i.e., 18 formula units of $CaSiO_3$, 90 atoms in total) separated by 15 Å vacuum spacing. These surface calculations employed a lower cutoff energy of 500 eV and a 1 × 3 × 3 k-point mesh. All atoms were free to relax. The geometries were optimized using the conjugate gradient method until the residual forces on the atoms were smaller than 0.01 eV/Å. The criterion for electronic convergence was set to an energy change of less than $10^{-6}$ eV. Differential adsorption energies ($\Delta E_{ads}$), defined as the energy required to add $\Delta n$ molecules to a surface already containing $n$ pre-adsorbed molecules, were calculated according to:

$$\Delta E_{ads} = \frac{E_{(n+\Delta n)} - E_n - \Delta n \cdot E_{H_2O}}{\Delta n}$$

where $E_{(n+\Delta n)}$ and $E_n$ were the relaxed total energies of the surface slab after and before the addition of $\Delta n$ molecules, respectively. $E_{H_2O}$ was the total energy of an isolated water molecule in the gas phase.

**Structure search strategy.** Water adsorption configurations were investigated at coverages between 1 and 5 $H_2O$ molecules per (1 × 1) unit cell. The dimensions of the computational supercells were selected to replicate the experimentally observed periodicities. Accordingly, (1 × 1) cells were employed to model "nested" (1 $H_2O$/u.c.) and "protruding" (2 $H_2O$/u.c.) water in the low-coverage regime. The "stable √2" (2.5 $H_2O$/u.c.) and (2 × 1) (4 $H_2O$/u.c.) patterns were simulated using (√2 × √2)R45° and (2 × 1) unit cells, respectively. Throughout this work, surface patterns are described using the standard surface science convention, which lists the scaling factor for the shorter surface lattice vector first. Consistent with the loss of long-range order in nc-AFM images at exposures exceeding 3 L (> 4 $H_2O$/u.c.), a standard (1 × 1) unit cell was adopted to investigate the high-coverage limit of 5 $H_2O$ molecules. To reduce computational cost for the structure search, the wollastonite substrate was modeled using a symmetric slab with a thickness of one unit bulk cell (i.e., 6 $CaSiO_3$ formula units, 30 atoms in total per (1 × 1) unit cell), thinner than the one used for obtaining the final optimized structure models and the calculated adsorption energies. The atoms in the lower half of this slab were frozen in the bulk positions. Before studying adsorption, an individual water molecule was optimized in the gas phase within an 8 Å cubic unit cell. Subsequently, the optimized molecules were placed in randomized starting positions to avoid a systematic bias. These placements were constrained such that all atoms of the generated water molecules were located at a height between 1.5 Å and 4.5 Å above the wollastonite surface. To avoid non-physical structures, which would cause large repulsive forces during the DFT relaxations, the minimum distance between all atoms (intermolecularly and between water molecules and the surface) was set to 1.5 Å. A two-stage relaxation strategy was employed to efficiently screen the potential energy surface and reach more favorable configurations. For a preliminary screening, approximately 500 randomly generated structures per water coverage were pre-relaxed using the GGA PBE-D3 exchange-correlation functional.[59] This functional was selected because it is computationally less demanding than metaGGA functionals, explicitly includes van der Waals interactions, and is a good compromise for modeling solid–water interfaces.[60] Following this preliminary screening, the lowest-energy candidate structures were selected for final and accurate relaxation using the more expensive metaGGA $r^2$SCAN+rVV10 functional and symmetric slabs with a thickness of three bulk unit cells (see the DFT-optimized structure files St1–St5 included in the Supplementary Information). Consistency checks were performed on some configurations to ensure the reliability of the PBE-D3 functional for the pre-relaxation step. All final adsorption energies and optimized structures were calculated at the $r^2$SCAN+rVV10 level of theory.

**AFM simulations.** Non-contact AFM images of the relaxed models were simulated using the Probe-Particle Model,[44,61,62] which includes the electrostatic potential above the surface (derived from the DFT calculation), Lennard-Jones potentials, and the elastic properties of the tip. Cu, CuOx, and CO tips were simulated using the following values of lateral and vertical spring constants and charges (Cu: $k_{x,y}$ = 0.75 N/m, $k_z$ = 50.7 N/m, effective tip charge –0.05 e; CuOx: $k_{x,y}$ = 161.9 N/m, $k_z$ = 271.1 N/m, effective tip charge –0.05 e; CO: $k_{x,y}$ = 1.7 N/m, $k_z$ = 326.9 N/m, effective tip charge –0.005 e). The experimental oscillation amplitude was used in the simulations. Since the exact height of the tip in the experiment is unknown, a height was chosen that yielded the best visual agreement between the experiment and the simulation.

## Conflicts of interest

There are no conflicts to declare.

## Acknowledgements

This research was funded by the European Research Council (ERC) under the European Union's Horizon 2020 research and innovation programme (grant agreement No. 883395, Advanced Research Grant 'WatFun') and by the Austrian Science Fund (FWF) (Elise Richter project 'SURREAL', 10.55776/RIC4539124). For open access purposes, the author has applied for a CC BY public copyright license to any author accepted manuscript version arising from this submission. The computational results have been achieved using the Austrian Scientific Computing (ASC) infrastructure. The help of Christian Lengauer and Dominik Talla (Uni Wien) during pXRD data acquisition and treatment is gratefully acknowledged.

## References

1. M. A. Henderson, *Surf. Sci. Rep.*, 2002, **46**, 1–308.
2. A. Verdaguer, G. M. Sacha, H. Bluhm and M. Salmeron, *Chem. Rev.*, 2006, **106**, 1478–1510.
3. H.-J. Freund and G. Pacchioni, *Chem. Soc. Rev.*, 2008, **37**, 2224–2242.
4. O. Björneholm, M. H. Hansen, A. Hodgson, L.-M. Liu, D. T. Limmer, A. Michaelides, P. Pedevilla, J. Rossmeisl, H. Shen, G. Tocci, E. Tyrode, M.-M. Walz, J. Werner and H. Bluhm, *Chem. Rev.*, 2016, **116**, 7698–7726.
5. I. G. Richardson, *Cem. Concr. Res.*, 2008, **38**, 137–158.
6. A. Michaelides and K. Morgenstern, *Nat. Mater.*, 2007, **6**, 597–601.
7. P. A. Thiel and T. E. Madey, *Surf. Sci. Rep.*, 1987, **7**, 211–385.
8. A. Hodgson and S. Haq, *Surf. Sci. Rep.*, 2009, **64**, 381–451.
9. J. Carrasco, A. Hodgson and A. Michaelides, *Nat. Mater.*, 2012, **11**, 667–674.
10. J. Matthiesen, S. Wendt, J. Ø. Hansen, G. K. H. Madsen, E. Lira, P. Galliker, E. K. Vestergaard, R. Schaub, E. Lægsgaard, B. Hammer and F. Besenbacher, *ACS Nano*, 2009, **3**, 517–526.
11. R. Włodarczyk, M. Sierka, K. Kwapień, J. Sauer, E. Carrasco, A. Aumer, J. F. Gomes, M. Sterrer and H.-J. Freund, *J. Phys. Chem. C*, 2011, **115**, 6764–6774.
12. X. Zhao, X. Shao, Y. Fujimori, S. Bhattacharya, L. M. Ghiringhelli, H.-J. Freund, M. Sterrer, N. Nilius and S. V. Levchenko, *J. Phys. Chem. Lett.*, 2015, **6**, 1204–1208.
13. Y. Fujimori, X. Zhao, X. Shao, S. V. Levchenko, N. Nilius, M. Sterrer and H.-J. Freund, *J. Phys. Chem. C*, 2016, **120**, 5565–5576.
14. O. Dulub, B. Meyer and U. Diebold, *Phys. Rev. Lett.*, 2005, **95**, 136101.

15. M. Wagner, P. Lackner, S. Seiler, A. Brunsch, R. Bliem, S. Gerhold, Z. Wang, J. Osiecki, K. Schulte, L. A. Boatner, M. Schmid, B. Meyer and U. Diebold, *ACS Nano*, 2017, **11**, 11531–11541.
16. M. Meier, J. Hulva, Z. Jakub, J. Pavelec, M. Setvin, R. Bliem, M. Schmid, U. Diebold, C. Franchini and G. S. Parkinson, *Proc. Natl. Acad. Sci. U.S.A.*, 2018, **115**, E5642–E5650.
17. X. M. Aretxabaleta, J. López-Zorrilla, I. Etxebarria and H. Manzano, *Nat. Commun.*, 2023, **14**, 7979.
18. R. J.-M. Pellenq, A. Kushima, R. Shahsavari, K. J. Van Vliet, M. J. Buehler, S. Yip and F.-J. Ulm, *Proc. Natl. Acad. Sci. U.S.A.*, 2009, **106**, 16102–16107.
19. Z. Casar, A. K. Mohamed, P. Bowen and K. Scrivener, *J. Phys. Chem. C*, 2023, **127**, 18652–18661.
20. N. Giraudo, P. Krolla-Sidenstein, S. Bergdolt, M. Heinle, H. Gliemann, F. Messerschmidt, P. Brüner and P. Thissen, *J. Phys. Chem. C*, 2015, **119**, 10493–10499.
21. T. K. Kundu, K. Hanumantha Rao and S. C. Parker, *Chem. Phys. Lett.*, 2003, **377**, 81–92.
22. T. K. Kundu, K. Hanumantha Rao and S. C. Parker, *J. Phys. Chem. B*, 2005, **109**, 11286–11295.
23. S. Sanna, W. G. Schmidt and P. Thissen, *J. Phys. Chem. C*, 2014, **118**, 8007–8013.
24. R. C. Longo, K. Cho, P. Brüner, A. Welle, A. Gerdes and P. Thissen, *ACS Appl. Mater. Interfaces*, 2015, **7**, 4706–4712.
25. P. Thissen, C. Natzeck, N. Giraudo, P. Weidler and C. Wöll, *Chem. - Eur. J.*, 2018, **24**, 8603–8608.
26. F. J. Giessibl, *Rev. Sci. Instrum.*, 2019, **90**, 011101.
27. L. Gross, F. Mohn, N. Moll, P. Liljeroth and G. Meyer, *Science*, 2009, **325**, 1110–1114.
28. B. Schulze Lammers, D. Yesilpinar, A. Timmer, Z. Hu, W. Ji, S. Amirjalayer, H. Fuchs and H. Mönig, *Nanoscale*, 2021, **13**, 13617–13623.
29. J. Peng, J. Guo, R. Ma and Y. Jiang, *Surf. Sci. Rep.*, 2022, **77**, 100549.
30. J. Peng, J. Guo, P. Hapala, D. Cao, R. Ma, B. Cheng, L. Xu, M. Ondráček, P. Jelínek, E. Wang and Y. Jiang, *Nat. Commun.*, 2018, **9**, 122.
31. P. Chen, Q. Xu, Z. Ding, Q. Chen, J. Xu, Z. Cheng, X. Qiu, B. Yuan, S. Meng and N. Yao, *Nat. Commun.*, 2023, **14**, 5813.
32. R. Ma, D. Cao, C. Zhu, Y. Tian, J. Peng, J. Guo, J. Chen, X.-Z. Li, J. S. Francisco, X. C. Zeng, L.-M. Xu, E.-G. Wang and Y. Jiang, *Nature*, 2020, **577**, 60–63.
33. A. Shiotari, Y. Sugimoto and H. Kamio, *Phys. Rev. Mater.*, 2019, **3**, 093001.
34. A. Shiotari and Y. Sugimoto, *Nat. Commun.*, 2017, **8**, 14313.
35. S. Lebègue, J. Harl, T. Gould, J. G. Ángyán, G. Kresse and J. F. Dobson, *Phys. Rev. Lett.*, 2010, **105**, 196401.
36. J. Klimeš, D. R. Bowler and A. Michaelides, *J. Phys.: Condens. Matter*, 2010, **22**, 022201.
37. J. Klimeš, D. R. Bowler and A. Michaelides, *Phys. Rev. B*, 2011, **83**, 195131.
38. M. J. Gillan, D. Alfè and A. Michaelides, *J. Chem. Phys.*, 2016, **144**, 130901.

39. J. Ning, M. Kothakonda, J. W. Furness, A. D. Kaplan, S. Ehlert, J. G. Brandenburg, J. P. Perdew and J. Sun, *Phys. Rev. B*, 2022, **106**, 075422.
40. H. Peng, Z.-H. Yang, J. P. Perdew and J. Sun, *Phys. Rev. X*, 2016, **6**, 041005.
41. P. Pedevilla, S. J. Cox, B. Slater and A. Michaelides, *J. Phys. Chem. C*, 2016, **120**, 6704–6713.
42. M. Ceriotti, W. Fang, P. G. Kusalik, R. H. McKenzie, A. Michaelides, M. A. Morales and T. E. Markland, *Chem. Rev.*, 2016, **116**, 7529–7550.
43. T. E. Markland and M. Ceriotti, *Nat. Rev. Chem.*, 2018, **2**, 0109.
44. N. Oinonen, A. V. Yakutovich, A. Gallardo, M. Ondráček, P. Hapala and O. Krejčí, *Comput. Phys. Commun.*, 2024, **305**, 109341.
45. A. Conti, L. Lezuo, A. Hoheneder, E. Vaníčková, D. A. Aloi, A. Steiger-Thirsfeld, D. Heuser, R. Abart, F. Mittendorfer, M. Schmid, U. Diebold and G. Franceschi, *ACS Nano, accepted*, 2026.
46. M. J. Buerger and C. T. Prewitt, *Proc. Natl. Acad. Sci. U.S.A.*, 1961, **47**, 1884–1888.
47. T. Ito, R. Sadanaga, Y. Takéuchi and M. Tokonami, *Proc. Jpn. Acad.*, 1969, **45**, 913–918.
48. G. Franceschi, A. Conti, L. Lezuo, R. Abart, F. Mittendorfer, M. Schmid and U. Diebold, *J. Phys. Chem. Lett.*, 2024, **15**, 15–22.
49. J. Balajka, J. Pavelec, M. Komora, M. Schmid and U. Diebold, *Rev. Sci. Instrum.*, 2018, **89**, 083906.
50. F. Huber and F. J. Giessibl, *Rev. Sci. Instrum.*, 2017, **88**, 073702.
51. S. Sadewasser and T. Glatzel, *Kelvin Probe Force Microscopy: Measuring and Compensating Electrostatic Forces*, Springer Berlin, Heidelberg, 2012.
52. C. A. Schneider, W. S. Rasband and K. W. Eliceiri, *Nat. Methods*, 2012, **9**, 671–675.
53. J. I. J. Choi, W. Mayr-Schmölzer, F. Mittendorfer, J. Redinger, U. Diebold and M. Schmid, *J. Phys.: Condens. Matter*, 2014, **26**, 225003.
54. R. Jinnouchi, J. Lahnsteiner, F. Karsai, G. Kresse and M. Bokdam, *Phys. Rev. Lett.*, 2019, **122**, 225701.
55. A. Omranpour, P. Montero De Hijes, J. Behler and C. Dellago, *J. Chem. Phys.*, 2024, **160**, 170901.
56. G. Kresse and D. Joubert, *Phys. Rev. B*, 1999, **59**, 1758–1775.
57. P. E. Blöchl, *Phys. Rev. B*, 1994, **50**, 17953–17979.
58. G. Kresse and J. Furthmüller, *Comput. Mater. Sci.*, 1996, **6**, 15–50.
59. S. Grimme, S. Ehrlich and L. Goerigk, *J. Comput. Chem.*, 2011, **32**, 1456–1465.
60. J. I. Hütner, A. Conti, D. Kugler, F. Sabath, K. N. Dreier, H.-G. Stammler, F. Mittendorfer, A. Kühnle, M. Schmid, U. Diebold and J. Balajka, *Sci. Adv.*, 2025, **11**, eaea2378.
61. P. Hapala, G. Kichin, C. Wagner, F. S. Tautz, R. Temirov and P. Jelínek, *Phys. Rev. B*, 2014, **90**, 085421.
62. P. Hapala, R. Temirov, F. S. Tautz and P. Jelínek, *Phys. Rev. Lett.*, 2014, **113**, 226101.
63. J. Heggemann, S. Aeschlimann, T. Dickbreder, Y. S. Ranawat, R. Bechstein, A. Kühnle, A. S. Foster and P. Rahe, *Phys. Chem. Chem. Phys.*, 2024, **26**, 21365–21369.
64. M. Wagner, B. Meyer, M. Setvin, M. Schmid and U. Diebold, *Nature*, 2021, **592**, 722–725.

Supporting Information for

# Water adsorption on a model silicate surface: wollastonite (100)

Luca Lezuo,[1†] Andrea Conti,[1†] Alexander Hoheneder,[1] Elena Vaníčková,[2] Domitilla Alessandra Aloi,[1] Rainer Abart,[3] Florian Mittendorfer,[1] Michael Schmid,[1] Ulrike Diebold,[1] Giada Franceschi[1*]

[1]Institute of Applied Physics, TU Wien, 1040, Vienna, Austria
[2]Central European Institute of Technology, Brno University of Technology, 61200, Brno, Czech Republic
[3]Department of Lithospheric Research, Universität Wien, 1090, Vienna, Austria
† These authors contributed equally.

*Corresponding author: Giada Franceschi
**Email:** franceschi@iap.tuwien.ac.at

**This PDF file includes:**

**Structure files (separate files):**

- St1: (1 × 1) nested $H_2O$ model (Fig. 1a). File name: "St1_1x1_nested.cif"
- St2: (1 × 1) protruding $H_2O$ model (Fig. 2a). File name: "St2_1x1_protruding.cif"
- St3: "stable √2" model (Fig. 4a). File name: "St3_stable_sqrt2.cif"
- St4: (2 × 1) model (Fig. 5f). File name: "St4_2x1.cif"
- St5: (1 × 1) model with 5$H_2O$/u.c. (Fig. S8). File name: "St5_1x1_5H2O.cif"

## Section S1: Additional nc-AFM data and AFM simulations

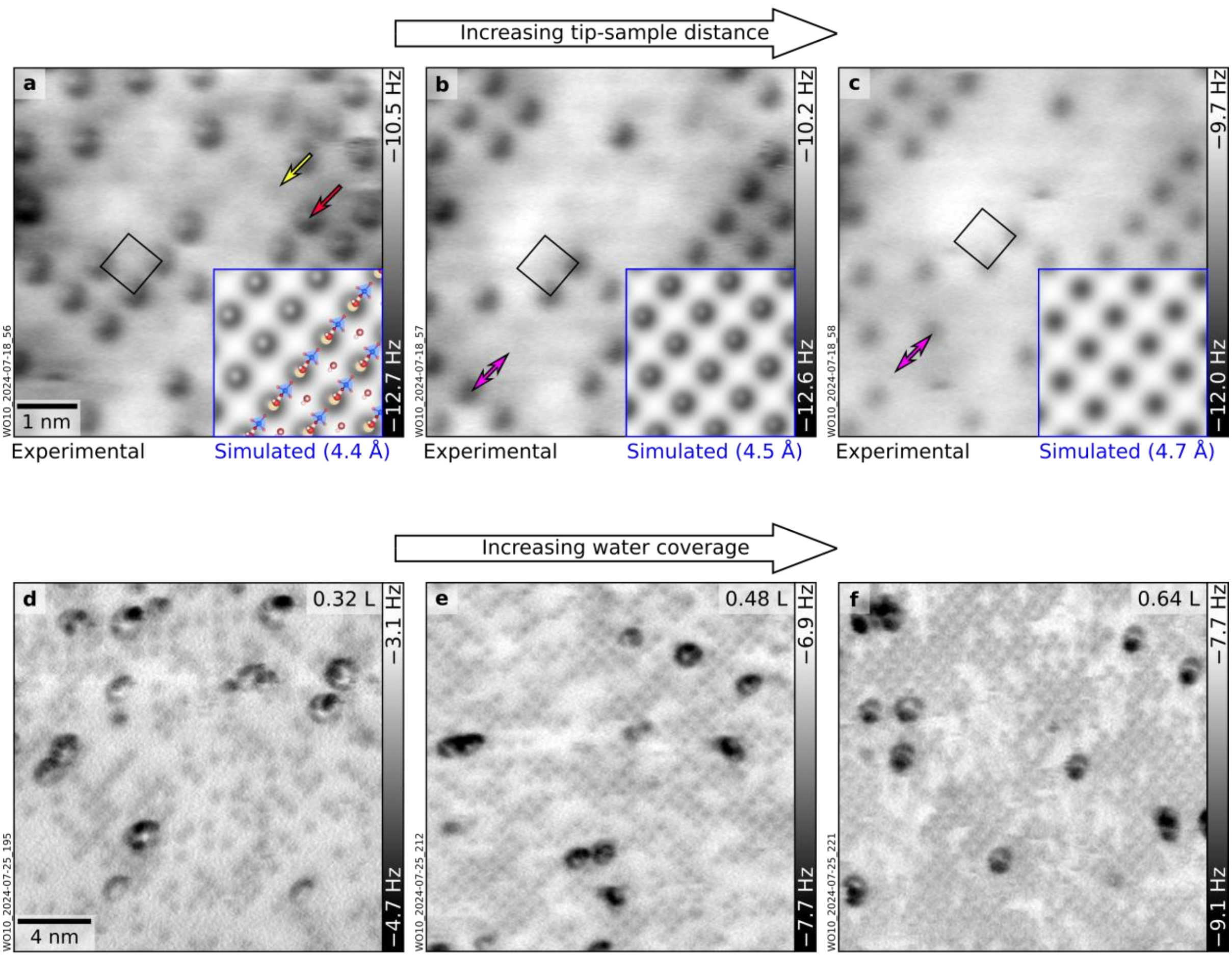

**Fig. S1: Protruding water molecules at different tip-sample distances and coverages.** (a–c) Consecutive nc-AFM images (6.1 × 6.1 nm$^2$) taken with increasing tip–sample distance (20 pm between each frame) on a surface partially covered with protruding water molecules. Surface Ca atoms are faintly visible (yellow arrow); the protruding water (red arrow) sits at equivalent lattice positions. Some molecules (magenta double arrows) undergo changes in position between consecutive images. Insets: AFM simulations based on the model shown in Figure 2a of the main text (tip–sample distances in brackets). (d–f) 20.0 × 20.0 nm$^2$ nc-AFM images of the cleaved surface after water exposure to (d) 0.32 L (resulting in a coverage of 0.33 ML), (e) 0.48 L (0.55 ML), and (f) 0.64 L (0.75 ML). AFM parameters: (a–c) $A$ = 200 pm, $V_s$ = –10 V; (d) $A$ = 540 pm, $V_s$ = –10 V; (e) $A$ = 580 pm, $V_s$ = –10 V; (f) $A$ = 480 pm, $V_s$ = –10 V.

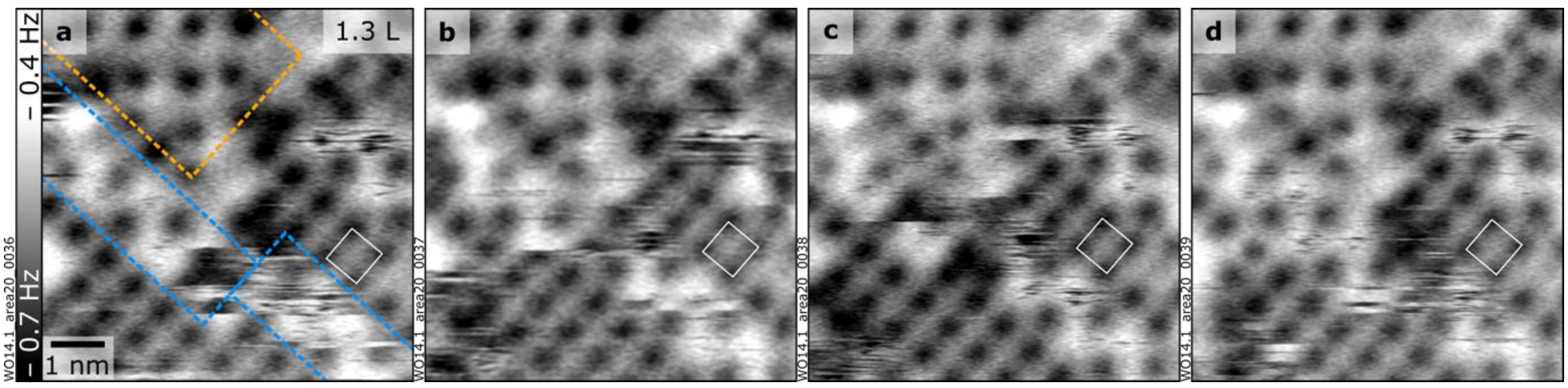

**Figure S2: Mobility of water species within the "stripes".** (a–d) Consecutive (Δt ≈ 12.5 min between successive images) nc-AFM images (7 × 7 nm$^2$) of wollastonite (100) covered with 2.3 $H_2O$ molecules; the white square marks the (1 × 1)-unit cell in the same spot. (a) Dashed blue rectangles indicate stripes, while the dashed orange rectangle indicates a stable √2 patch. $A$ = 700 pm, $V_s$ = –10.0 V, O-terminated tip.

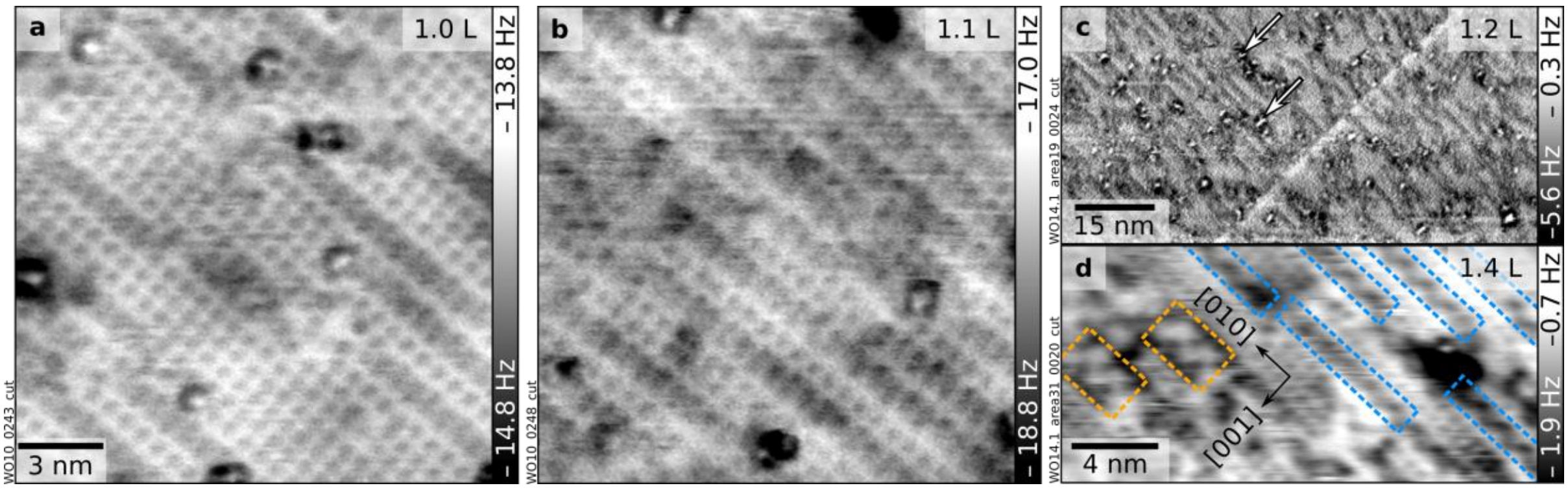

**Figure S3: Increasing density of stripes.** (a–d) nc-AFM images of wollastonite (100) with increasing coverages of water (doses in Langmuir, in addition to the nested water, are indicated at the top right corner of each panel). (a, b) 17 × 17 nm$^2$ images showing the increasing density of the stripes with increasing coverage (the stable √2 patches are not resolved with this tip). (c) 90 × 45 nm$^2$ image: The stripes often end at surface defects (arrows). (d) 22 × 11 nm$^2$ image highlighting the coexistence of stripes (blue markings) and √2 patches (orange markings), which can be distinguished with this tip. AFM parameters: (a) $A$ = 490 pm, $V_s$ = –10.0 V; (b) $A$ = 400 pm, $V_s$ = –10.0 V; (c) $A$ = 700 pm, $V_s$ = –8.5 V; (d) $A$ = 480 pm, $V_s$ = –3.5 V, Cu-terminated tip.

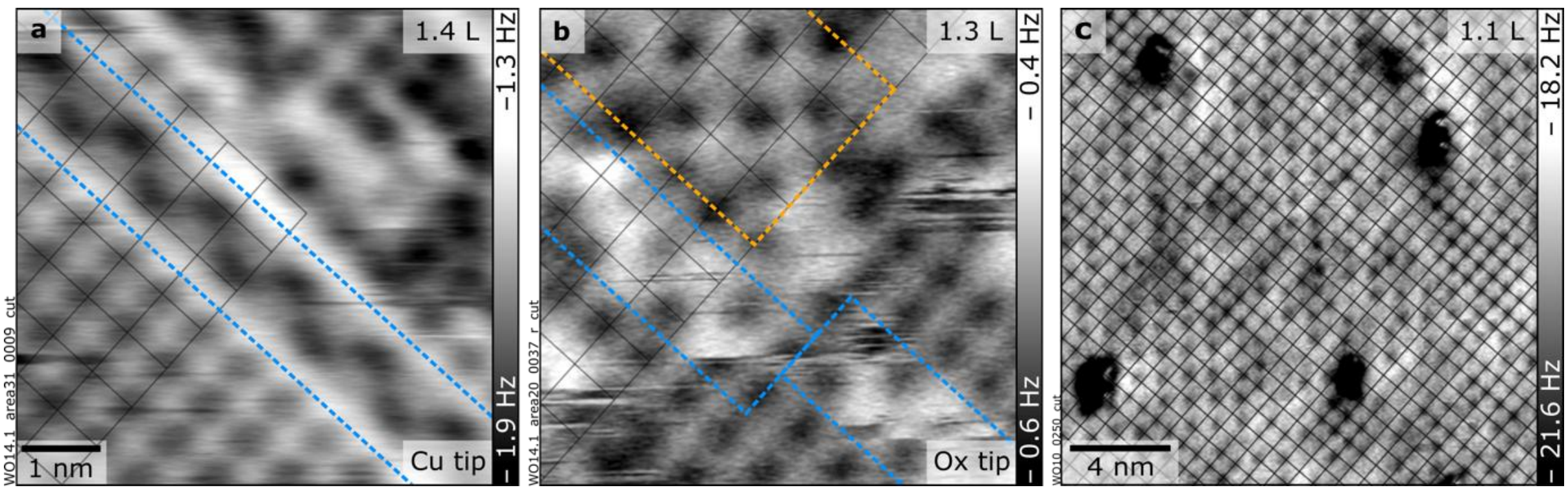

**Figure S4: Relative alignment of different water structures.** (a, b) 6.5 × 6.5 nm$^2$ and (c) 19.0 × 19.0 nm$^2$ nc-AFM images of wollastonite (100) plus water at a coverage of 2.1–2.4 $H_2O$ molecules per (1 × 1) unit cell (doses in Langmuir, in addition to the nested water, are reported at the top right corners of the corresponding panels). The (1 × 1) unit cell (black grid) is shifted with respect to the dark spots with (√2 × √2)R45° periodicity; note that the magnitude of the apparent shift depends on the tip termination. Stripes and stable √2 patches are highlighted in blue and orange in panels (a) and (b), respectively; the tip in panel (c) does not distinguish between the (1 × 1) areas and the stripes. AFM parameters: (a) $A$ = 700 pm, $V_s$ = –3.5 V, Cu-terminated tip; (b) $A$ = 700 pm, $V_s$ = –10.0 V, O-terminated tip; (c) $A$ = 350 pm, $V_s$ = –10.0 V, wollastonite-modified tip.

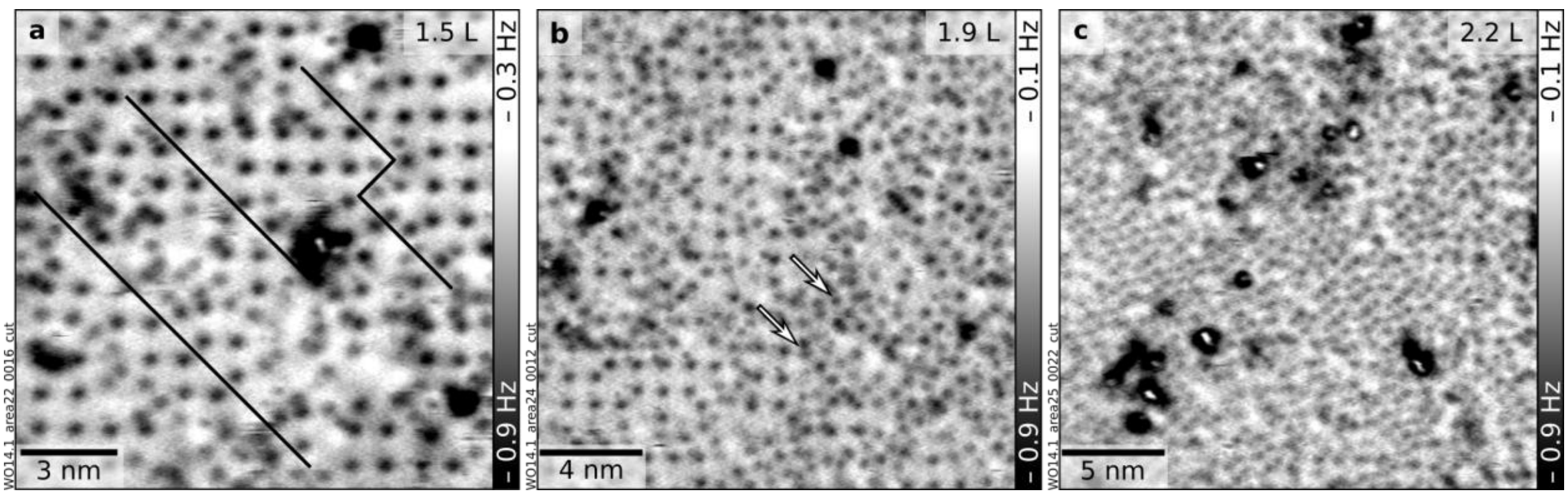

**Figure S5: More details about the transition between the "stable √2" and the (2 × 1) structure.** (a–c) 15 × 15, 18.7 × 18.7, and 23.3 × 23.3 $nm^2$ nc-AFM images of coverage between 2.5 and 3.2 $H_2O$ molecules per (1 × 1) unit cell (doses in Langmuir, in addition to the nested water, are indicated at the top right corners). (a) Slightly beyond the full "stable √2" pattern: Black lines highlight domain boundaries. (b) A preferential ordering along the [010] direction becomes apparent (see arrows). (c) Nearly complete (2 × 1) layer. AFM parameters: (a) $A$ = 700 pm, $V_s$ = –2.7 V; (b) $A$ = 700 pm, $V_s$ = –3.6 V; (c) $A$ = 700 pm, $V_s$ = –2.5 V; O-terminated tips.

## Section S2: Additional computational results

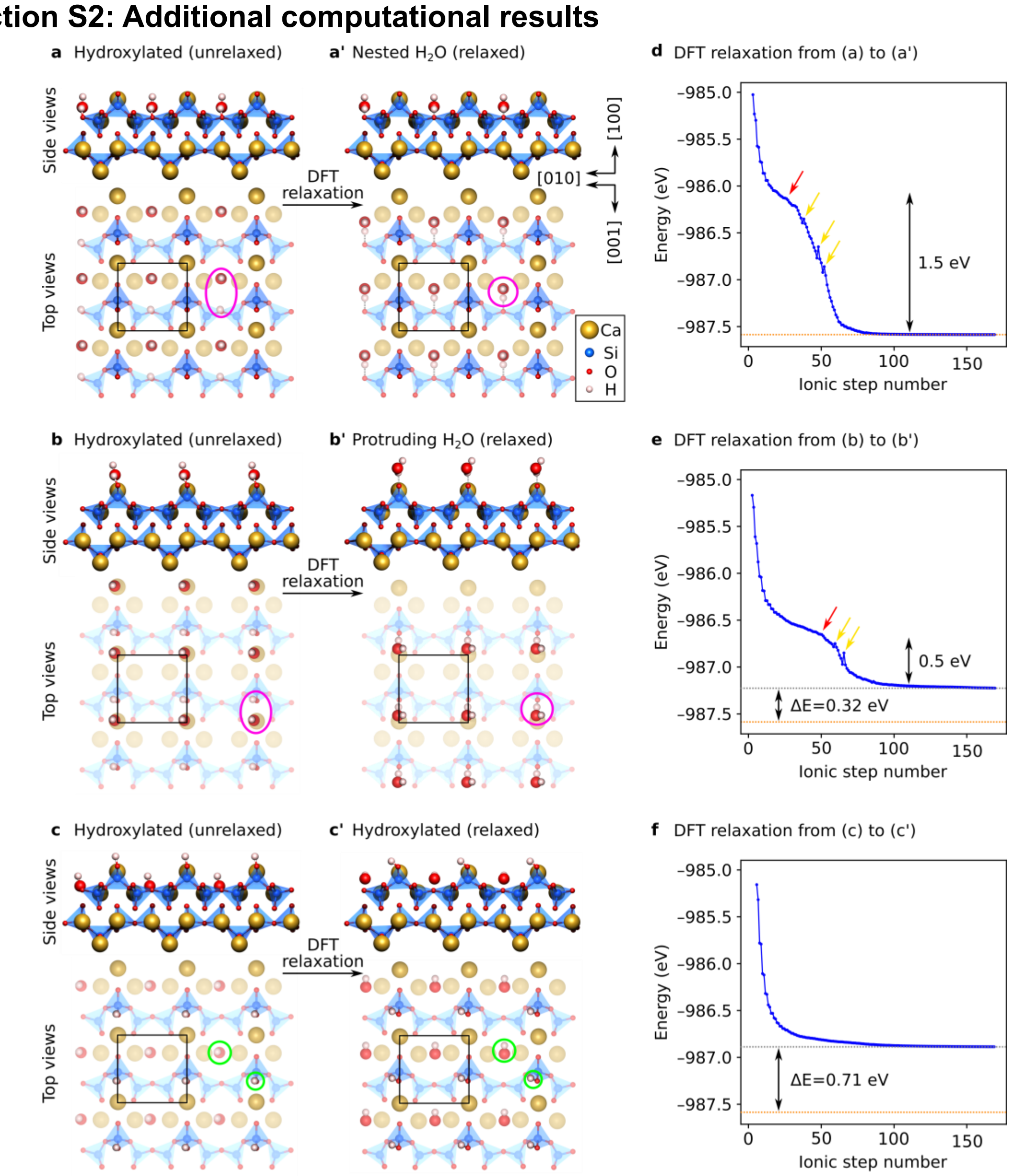

**Figure S6: Energetic preference for molecular water adsorption on the pristine wollastonite surface.** (a–c) Initial unrelaxed configurations representing hypothetical dissociative adsorption states, with hydroxyl groups and protons coordinated to surface Ca ions and bridging or apical oxygens of silica tetrahedra, respectively. (a'–c') Final relaxed configurations obtained after DFT optimization at 0 K. Dissociated species spontaneously recombine into intact water molecules in (a') nested or (b') protruding configurations (magenta circles); (c') widely separated species remain dissociated (green circles). (d–f) Energy profiles along the relaxation path from the dissociated states (a–c) to the molecular states (a'–b') and hydroxylated state (c'). Transient energy spikes (yellow arrows) reflect conjugate gradient trial steps probing repulsive regions during O–H bond formation. Red arrows indicate the onset of molecular recombination. The nested molecular state is the most stable (dotted orange line), while protruding molecular and fully hydroxylated states are 0.32 eV/$H_2O$ and 0.71 eV/$H_2O$ less favorable, respectively. The significant energy gain upon water recombination (0.5–1.5 eV) demonstrates that molecular adsorption is favored over dissociation, in contrast to other silicate surfaces[1] or wollastonite (001).[2,3]

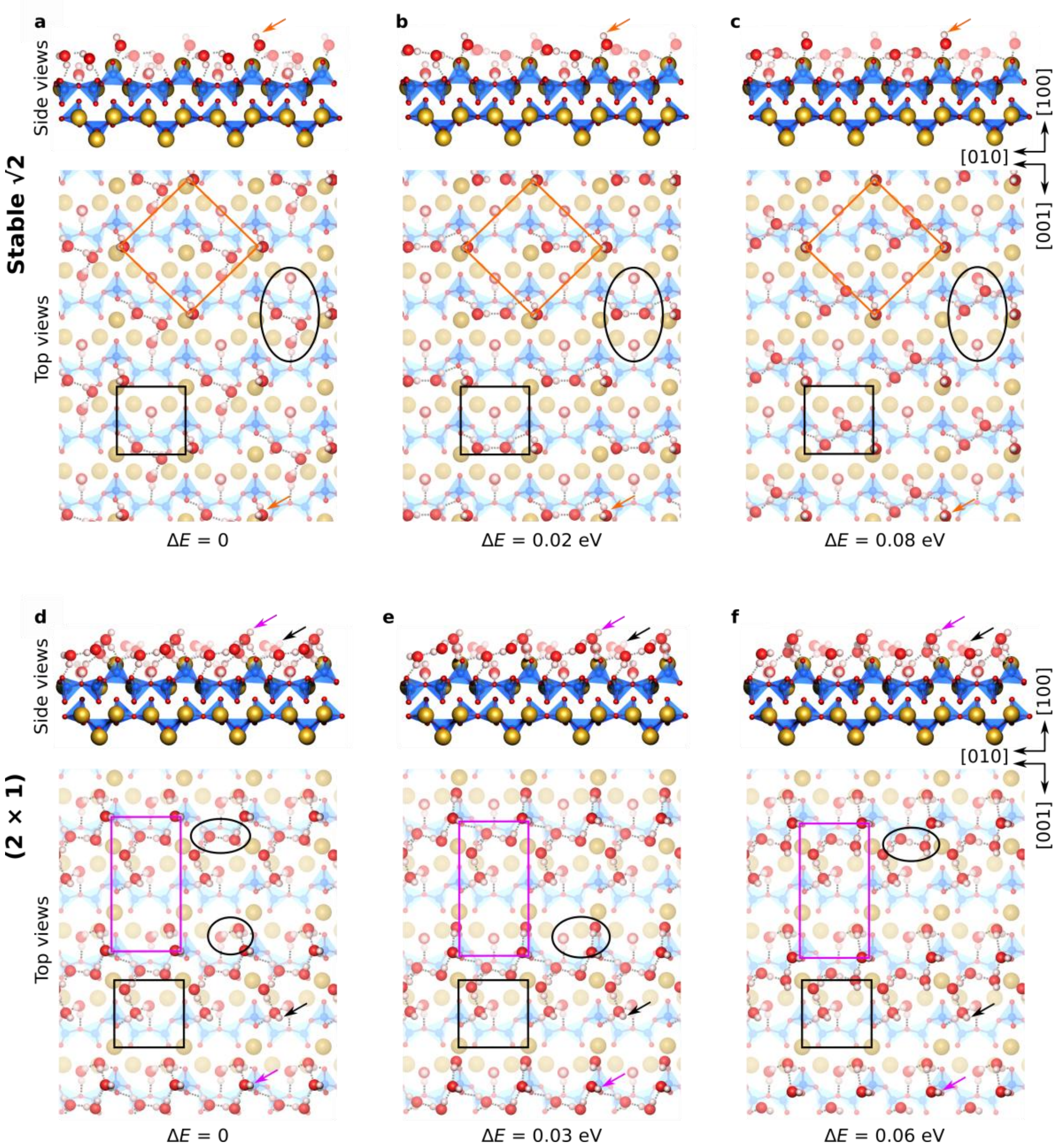

**Figure S7: Comparison of energetically similar water configurations for the "stable √2" and (2 × 1) structures.** (a) Lowest-energy model and (b, c) two low-energy models for a coverage of 2.5 $H_2O$ per (1 × 1) unit cell (black rectangle) in a (√2 × √2)R45° cell (orange rectangle), aiming to model the "stable √2" pattern observed experimentally. Orange arrows indicate the most protruding $H_2O$ molecule. (d) Lowest-energy model and (e, f) two low-energy models for a coverage of 4 $H_2O$ per (1 × 1) unit cell in a (2 × 1) cell (magenta rectangle). Magenta and black arrows indicate the most protruding and the second-most protruding $H_2O$ molecules, respectively. Black ovals mark molecules arranged in distinct configurations among the different models. Because these molecules do not protrude from the surface, they remain undetectable in constant-height nc-AFM measurements. Relative energy differences per unit cell (Δ*E*) are shown below the corresponding models. Since these differences (<0.1 eV per unit cell) lie within the typical accuracy limits of DFT for H-bonded systems,[4] these configurations are degenerate within the error bars and may coexist in experiments at finite temperatures.

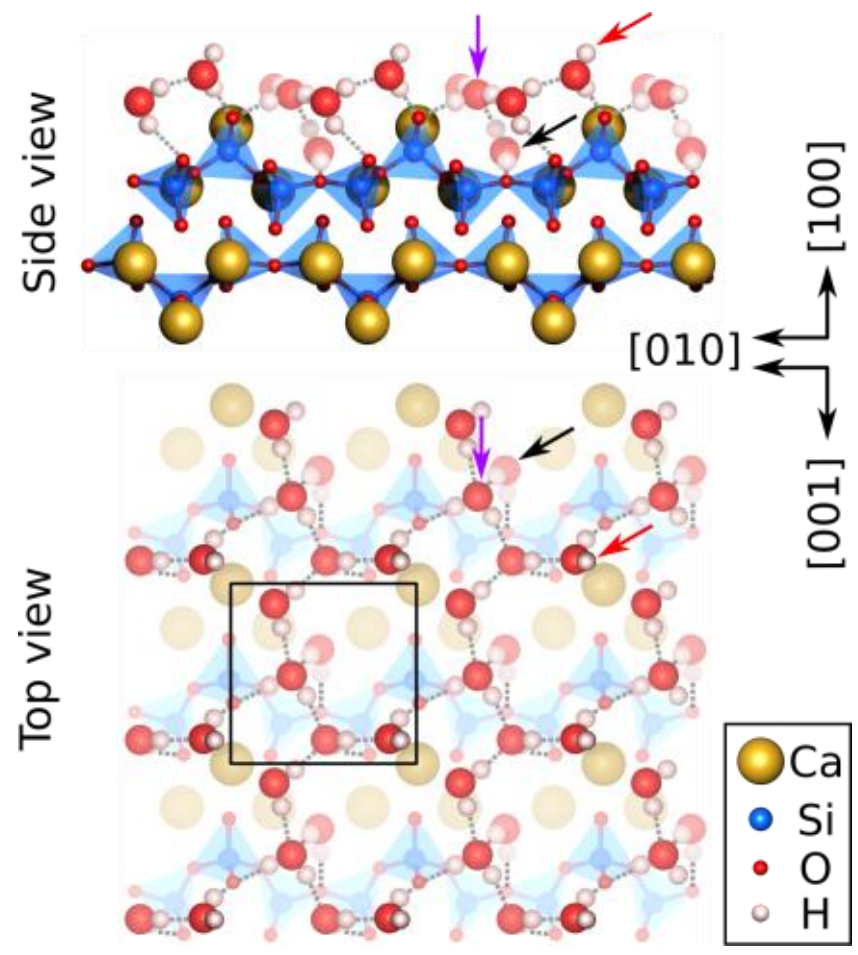

**Figure S8: Lowest-energy model with a coverage of five $H_2O$ per (1 × 1) unit cell.** Water molecules are arranged into an extended H-bonded network (marked by gray dotted lines). Although the initial adsorption geometries were generated through unbiased random sampling, the most favorable model exhibits one protruding $H_2O$ (red arrow) per unit cell bound to the Ca and H-bound to the apical oxygen of a surface silica tetrahedron. This suggests that this surface site remains active during water adsorption for all coverages where not all $H_2O$ can be accommodated as nested water. In contrast to structures at lower coverage, a second $H_2O$ (purple arrow) donates a H-bond to the same apical oxygen. This second molecule adopts a horizontal orientation and fourfold coordination: donating two H-bonds (to the apical surface oxygen and a neighboring $H_2O$) and accepting two (from a planar $H_2O$ and the nested $H_2O$). Similar to the (2 × 1) structure (see Figure 5f), the nested water (black arrow) is involved in the H-bonded network and remains strongly bound to its site, indicating a preference for this configuration even at the high-coverage limit. The differential adsorption energy per water molecule is reported in Figure 5d.

| $CaSiO_3$ | | a (rel. error) | b (rel. error) | c (rel. error) | α (rel. error) | β (rel. error) | γ (rel. error) | Cell volume (rel. error) | Bulk modulus (rel. error) | Band gap (rel. error) |
|---|---|---|---|---|---|---|---|---|---|---|
| **Experiment (RT)** | | 7.93 Å | 7.32 Å | 7.07 Å | 90.1° | 95.2° | 103.4° | 396.9 Å$^3$ | 102.0 GPa | 5.45 eV |
| **GGA** | **PBE** | 8.03 Å (+1.3%) | 7.40 Å (+1.1%) | 7.16 Å (+1.3%) | 90.1° (+0.0%) | 95.5° (+0.3%) | 103.5° (+0.0%) | 411.5 Å$^3$ (+3.7%) | 91.4 GPa (−10.4%) | 4.86 eV (−10.9%) |
| | **PBE-D3** | 7.95 Å (+0.4%) | 7.35 Å (+0.4%) | 7.09 Å (+0.4%) | 90.0° (+0.0%) | 95.4° (+0.1%) | 103.4° (+0.0%) | 401.3 Å$^3$ (+1.1%) | 93.8 GPa (−8.1%) | 4.99 eV (−8.5%) |
| **metaGGA** | **$r^2$SCAN** | 7.92 Å (−0.1%) | 7.31 Å (−0.1%) | 7.07 Å (+0.1%) | 90.0° (+0.0%) | 95.3° (+0.1%) | 103.4° (+0.0%) | 396.6 Å$^3$ (−0.1%) | 98.8 GPa (−3.1%) | 5.60 eV (+2.8%) |
| | **$r^2$SCAN-D3** | 7.88 Å (−0.5%) | 7.29 Å (−0.4%) | 7.04 Å (−0.3%) | 90.0° (+0.0%) | 95.2° (+0.0%) | 103.4° (+0.0%) | 392.2 Å$^3$ (−1.2%) | 100.8 GPa (−1.2%) | 5.65 eV (+3.7%) |
| | **$r^2$SCAN+rVV10** | 7.89 Å (−0.5%) | 7.30 Å (−0.2%) | 7.04 Å (−0.3%) | 90.0° (+0.0%) | 95.1° (−0.1%) | 103.4° (+0.0%) | 393.1 Å$^3$ (−1.0%) | 100.8 GPa (−1.2%) | 5.54 eV (+1.6%) |

**Table T1: Experimental and calculated bulk properties of wollastonite.** Lattice parameters (a, b, c, α, β, γ), unit cell volume, bulk modulus, and band gap obtained for different exchange–correlation functionals compared with experimental data. Experimental unit cell parameters were derived from XRD analysis[5] at room temperature (RT), while the bulk modulus and band gap values were taken from Refs.[6,7], respectively. Brackets indicate relative percentage errors with respect to experimental reference. The PBE-D3 and $r^2$SCAN+rVV10 functionals provide the most accurate description of the bulk phase (lattice parameters, bulk modulus, and band gap) within the GGA and metaGGA families, respectively.

| System | Surface unit cell dimensions | Surface unit cell area | Water coverage per surface unit cell | Surface water density | Experimentally observed structures |
|---|---|---|---|---|---|
| **Bulk ice Ih** | 7.79 Å × 7.79 Å | 52.6 $Å^2$ | 6 $H_2O$ | 11.4 $H_2O/nm^2$ | - |
| **Silver iodide (AgI)** | 9.18 Å × 9.18 Å | 84.3 $Å^2$ | 8 $H_2O$ | 9.5 $H_2O/nm^2$ | Epitaxial ice Ih layer |
| **Wollastonite ($CaSiO_3$)** | 7.07 Å × 7.32 Å | 51.8 $Å^2$ | 2.5 $H_2O$ | 4.8 $H_2O/nm^2$ | Stable √2 |
| | | | 4 $H_2O$ | 7.7 $H_2O/nm^2$ | (2 × 1) |
| | | | 5 $H_2O$ | 9.7 $H_2O/nm^2$ | 3D clusters |

**Table T2: Surface water density for bulk ice Ih, silver iodide, and wollastonite.** Comparison of surface unit cell dimensions, areas, and corresponding surface water densities for bulk ice Ih,[8] (2 × 2)-reconstructed AgI(0001) (a perfect ice nucleator), and (100)-oriented wollastonite ($CaSiO_3$). Silver iodide promotes the growth of an epitaxial ice layer (9.5 $H_2O/nm^2$).[9] Conversely, while a coverage of 5 $H_2O$ per unit cell on wollastonite yields a very close surface water density (9.7 $H_2O/nm^2$), the rectangular substrate geometry precludes planar epitaxy, driving the formation of 3D clusters.

## References

1. G. Franceschi, A. Conti, L. Lezuo, R. Abart, F. Mittendorfer, M. Schmid and U. Diebold, *J. Phys. Chem. Lett.*, 2024, **15**, 15–22.
2. R. C. Longo, K. Cho, P. Brüner, A. Welle, A. Gerdes and P. Thissen, *ACS Appl. Mater. Interfaces*, 2015, **7**, 4706–4712.
3. P. Thissen, C. Natzeck, N. Giraudo, P. Weidler and C. Wöll, *Chem. - Eur. J.*, 2018, **24**, 8603–8608.
4. M. J. Gillan, D. Alfè and A. Michaelides, *J. Chem. Phys.*, 2016, **144**, 130901.
5. A. Conti, L. Lezuo, A. Hoheneder, E. Vaníčková, D. A. Aloi, A. Steiger-Thirsfeld, D. Heuser, R. Abart, F. Mittendorfer, M. Schmid, U. Diebold and G. Franceschi, *ACS Nano, accepted*, 2026.
6. N. I. Demidenko and A. P. Stetsovskii, *Glass Ceram.*, 2003, **60**, 217–218.
7. H. Nagabhushana, B. M. Nagabhushana, M. Kumar, H. B. Premkumar, C. Shivakumara and R. P. S. Chakradhar, *Philos. Mag.*, 2010, **90**, 3567–3579.
8. K. Röttger, A. Endriss, J. Ihringer, S. Doyle and W. F. Kuhs, *Acta Cryst. B*, 1994, **50**, 644–648.
9. J. I. Hütner, A. Conti, D. Kugler, F. Sabath, K. N. Dreier, H.-G. Stammler, F. Mittendorfer, A. Kühnle, M. Schmid, U. Diebold and J. Balajka, *Sci. Adv.*, 2025, **11**, eaea2378.